# Physics background at ILC at 500GeV and 1TeV

**M. Pandurović and I. Božović-Jelisavčić**

*"Vinča" Institute of Nuclear Sciences,*
*Mihajla Petrovića Alasa 12-14, Beograd, Srbija*
*E-mail*: milap@vinca.rs

ABSTRACT: Measurement of the integrated luminosity at the International Linear Collider (ILC) will be accomplished by counting the rate of small angle Bhabha scattering events. The physics requirements for ILC set the constraint on the relative precision of the luminosity measurement to be of a permille order. The required precision can be achieved by construction of a fine granulated electromagnetic calorimeter of high energy and polar angle resolution and by sufficient experimental control of numerous systematic effects. One of the leading systematic effects in luminosity measurement is the background originating from four-fermion processes, referred to as the physics background. In this paper a possible selection strategy to measure the luminosity is proposed from the perspective of optimal signal to background separation.

___________________________________________________________________

## Contents

## 1. Introduction

Integrated luminosity at ILC will be determined from the total number of Bhabha events $N_{bha}$ produced in the fiducial volume of the luminosity calorimeter in the certain time interval and from the corresponding theoretical cross-section $\sigma_B$.

$$L_{int} = \frac{N_B}{\sigma_B} \qquad (1)$$

However, due to the presence of background events $N_{bck}$ that are misidentified as a signal, the number of counted Bhabha events $N_{exp}$ has to be corrected for miscounts. In addition, integral luminosity also has to be scaled by the selection efficiency $\varepsilon$:

$$L_{int} = \frac{N_{exp} - N_{bck}}{\varepsilon \cdot \sigma_B} \qquad (2)$$

Of course, other systematic corrections should be taken into account in the similar manner.

Bhabha scattering at small angles at the ILC energies, represent almost pure (99%) electromagnetic process of one-photon exchange in the t-channel, which is calculable with the great precision within QED. The theoretical uncertainty of the cross section of Bhabha scattering at the LEP energies is of $10^{-4}$ order [1]. Integrated Bhabha cross-section of the order of a few nb in the luminometer fiducial volume leads to the statistical error of the annual integrated luminosity less than $10^{-4}$.

At small angles the differential Bhabha scattering cross-section can be approximated by:

$$\frac{d\sigma^{QED}}{d\theta} \approx 32\pi\alpha^2 \frac{1}{s} \cdot \frac{1}{\theta^3} \qquad (3)$$

where s is the center-of-mass energy squared and θ is the polar angle of the final state particles. The strong dependence of the cross section of Bhabha scattering on the polar angle makes the luminosity measurement especially sensitive to the lower aperture of the fiducial volume of the luminometer.

Luminosity calorimeter (LumiCal) has been designed as a precision device for the integrated luminosity measurement. It is a compact electromagnetic sandwich calorimeter consisting of 30 longitudinal layers of silicon sensors followed by one radiation length of tungsten absorber and the interconnection structure. In the ILD concept [2], LumiCal is located at *z = 2500 mm* from the IP, covering the polar angle range between *31* mrad and *78* mrad. This design corresponds to the inner radius of the LumiCal of $r_{min}$=*80 mm*, and outer radius of $r_{max}$=195 mm. Sensors are segmented into pads with 48 azimuthal and 96 radial divisions. The proposed design results in a small Moliere radius of 1.5 cm and in energy and polar angle resolutions of $\frac{\sigma_E}{E} \approx \frac{\alpha}{\sqrt{E_{beam}}}, \alpha = 0.21\sqrt{GeV}$ and $\delta_\theta = (2.20 \pm 0.01) \cdot 10^{-2}$ mrad, respectively [2]. Stability of the sampling term α (Figure 1) [2] defines detector fiducial volume in the range [41,67] mrad. Event selection discussed in this paper assumes only events counted in the detector fiducial volume, where showers are fully contained.

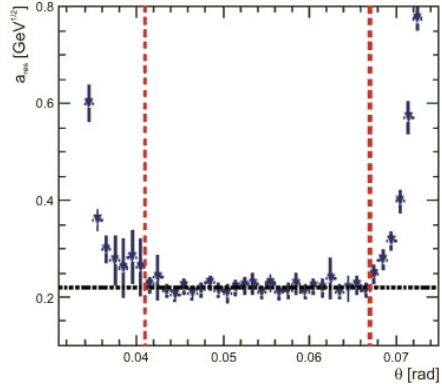

**Figure 1. The distribution of the relative energy resolution as a function of the polar angle.**

## 2. Event simulation

To estimate the background to signal ratio (B/S) relevant physics processes were simulated in the following way:

Two samples of 20 pb$^{-1}$ of Bhabha events, with cross-sections in the fiducial volume of $\sigma_B$=(4.689± 0.001) nb and $\sigma_B$=(1.197±0.005) nb, at 500GeV and 1TeV case respectively, are simulated, with BHLUMI [3] event generator, implemented in BARBIE V5.0 [4], a GEANT3 [5] based detector simulation of LumiCal.

Leptonic ($e^+e^- \to e^+e^-l\bar{l}$, l=e,μ,) and hadronic ($e^+e^- \to e^+e^-q\bar{q}$, q=u,d,s,c,b) events have been generated at 500 GeV and 1 TeV center-of-mass energy, with the total corresponding cross-section of *(5.1±0.1) nb* and *(0.77±0.11) nb*, respectively, using WHIZARD-V1.2 event generator [6]. Events were generated within the polar angle range [0.05,179.95] deg to avoid the divergency of the cross-section at low angles. The cross section integrated in the polar angle range of the fiducial volume is around percent of the generated one, for both center-of-mass energies. Matrix elements of the leading order Feynman diagrams are computed using O'Mega [7] generator. The total size of samples corresponds approximately to 1 million background events, generated at both ILC energies.

## 3. Event selection

Event selection for the luminosity measurement is, to some extent, based on isolation cuts as at LEP [1]: Bhabha scattering, in the Born approximation, being two-body elastic scattering is characterized by the back-to-back topology which implies colinearity and coplanarity of the signal tracks and equal energy deposits in the left and right side of the luminosity calorimeter. Strictly speaking, Born-level elastic Bhabha scattering never occurs. In practice, the process is always accompanied by emission of electromagnetic radiation. Emission of the initial and final state radiation and moreover beam-beam interaction effects [8] distort the primary signal signature affecting both signal topology and energy depositions. Therefore the selection criteria have to be adopted to accommodate for distortion of the signal signature, while efficiently eliminating background. The beam-beam effects will be briefly described in section 5. The detailed elaboration of the effect at ILC energies is given in [9][10][11].

The beam induced effects - beamstrahlung and electromagnetic deflection, are dominantly affecting the colinearity of the final state particles and energy loss of the particles thus resulting in the reduction of the center-of-mass energy. Therefore, high energy Bhabha pairs are selected by requiring that they carry more than 80% of the centre of mass energy and by allowing the limited acoplanarity of Bhabha particles up to a predefined limit.

Alternatively, energy balance $E_{bal}$ can be defined as a difference in energy deposits in the opposite arms of the detector:

$$E_{rel} = \frac{E_L + E_R}{2 \cdot E_{beam}} \qquad E_{bal} = |E_L - E_R| \qquad (4)$$

where $E_L$, $E_R$ stand for the particle energy deposited at the left and right side of the LumiCal.

In addition, an alternative selection based on based on LEP-like isolation cuts will be discussed to illustrate the limits of physics background suppression at the ILC energies.

## 4. Four-fermion background

The potential problem in luminosity measurement are events of the type $e^+e^- \rightarrow e^+e^- f\bar{f}$, since these processes have Bhabha like signature: outgoing pairs are emitted very close to the beam pipe carrying the most of the beam energy so they can be miscounted as a Bhabha pair. The leading order Feynman diagrams are given in Figure 2. The main contribution to the cross section is coming from the multiperipheral (two-photon exchange) processes.

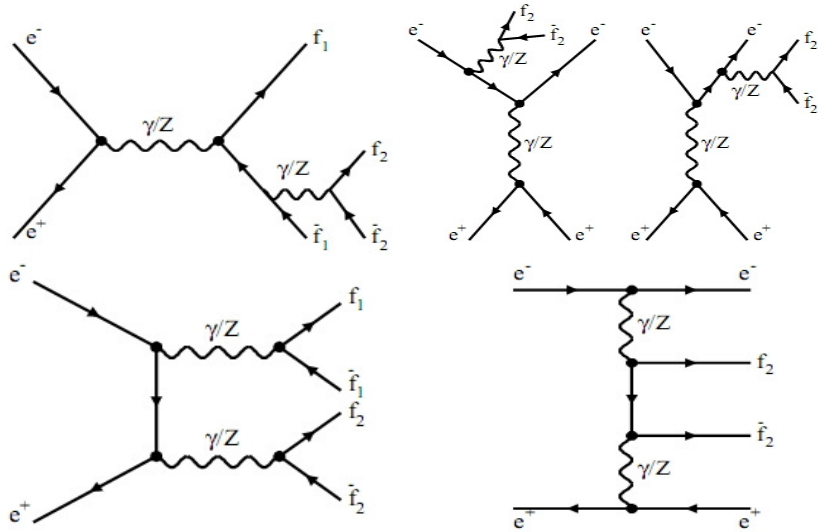

**Figure 2. The dominant Feynman diagrams for the neutral current four-fermion production: a) annihilation, b) bremsstrahlung, c) conversion and d) multiperipheral (two-photon).**

Due to the steep polar angle distribution of the electron spectators, the most of these particles are detected in the beam calorimeter, while a few permille are deposited in the luminosity calorimeter. Energy and polar angle distributions of hadronic and leptonic background in the acceptance of the luminosity calorimeter are given in Figure 3. As clearly visible in Figure 3a, electron spectators deposit maximal energy in the LumiCal peaking at the energy of a beam. Physics background polar angle distribution is similar to the signal one (Figure 3b) indicating very forward processes.

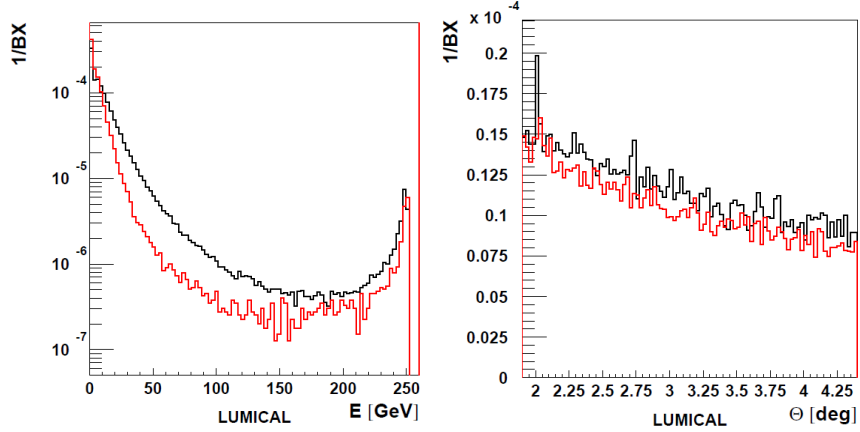

**Figure 3. Energy distributions (a) and polar angle distribution (b) in the acceptance of the luminosity calorimeter, for hadronic (black) and leptonic (red) physics background.**

Cross section of the two-photon processes is rising with the centre-of-mass energy as $\sim \ln^2(s)$ [12] and practically saturates at 10 nb in the vicinity of the ILC energies.

Parameters of the Monte Carlo event generator are tuned to the hadronic background ($e^+e^- \to e^+e^- c\bar{c}$) to describe the experimental results obtained by several experiments at PETRA and LEP colliders [13]. The results are given in Figure 4., along with the superimposed simulated cross section obtained by WHIZARD [6] and BDK [14] generators, under different assumptions on transferred momentum Q[GeV].

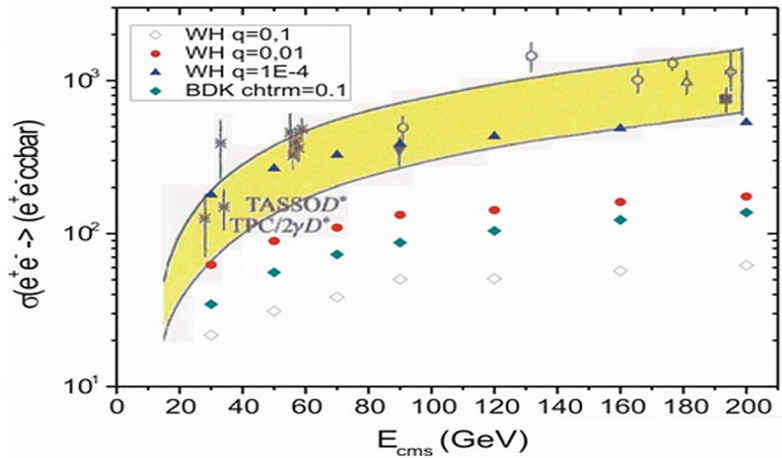

**Figure 4. Experimental results for the cross-section of $e^+e^- \to e^+e^- c\bar{c}$. The band corresponds to the NLO theoretical uncertainty. Event generation assumptions in WHIZARD and BDK are tuned to reproduce results at lower centre of mass energies.**

## 5. Beam-beam interaction effects

We take into account the fact that the beam induced effects are becoming more prominent with the rising center-of-mass energy (i.e. ILC w.r.t. LEP). Beamstrahlung prior and electromagnetic deflection after the Bhabha scattering result in the distortion of the measured luminosity spectrum as well as in the effective loss of Bhabha counts in the detector fiducial volume. Corrective methods, described in detail in [9] have been developed in order to correct for counting losses due to the beamstrahlung and the initial state radiation.

Applying the event selection as described in 3, it has been shown [9] that the contribution of these effects to the relative uncertainty of the measured luminosity is 0.5 permille at 500 GeV and 0.2 permille at 1 TeV center-of-mass energy.

## 6. Discussion

There are several variables to describe separation power of the proposed event selection: signal efficiency $E_s$, background rejection $R_{bck}$ and background to signal ratio B/S, where the uncertainty of the latter acctually contributes to the overall systematic uncertainty of the luminosity measurement.

$$E_s = \frac{N'_s}{N_s} \qquad R_{bck} = \frac{N'_{bck}}{N_{bck}} \qquad (5)$$

$$\frac{B}{S} = \frac{\sum \frac{N'_{bck}}{N^{gen}_{bck}/\sigma_{bck}}}{\frac{N'_s}{N^{gen}_s/\sigma_s}} \qquad (6)$$

where the $N_s$, $N_{bck}$ correspond to the number of coincidently detected pairs in the opposite sides of LumiCal for signal and background, respectively and the prime values correspond to the selected entries, $N^{gen}_s, N^{gen}_{bck}$ are numbers of signal and background events, respectively and $\sigma_s$, $\sigma_{bck}$ are corresponding cross sections. Since NLO corrections for the four-fermion production cross section are not yet known at the ILC energies, we will assume assume the physics background as a full size effect.

The proposed cut-off values are set to 0.8 for the fraction of the center of mass energy carried by a pair of final state particles and 5° for the azimuthal angle difference. Results are given in Table 1. for leptonic and hadronic background at 500 GeV and 1 TeV centre of mass energies.

|  |  | **500 GeV** | **1 TeV** |
|---|---|---|---|
| Signal | $E_s$ | 94% | 94% |
| Leptonic | $R_{bck}$ | 60% | 56% |
| background | B/S | $1.6 \cdot 10^{-3}$ | $0.7 \cdot 10^{-3}$ |
| Hadronic | $R_{bck}$ | 70% | 91% |
| Background | B/S | $0.6 \cdot 10^{-3}$ | $0.1 \cdot 10^{-3}$ |
| **Total** | **B/S** | $2.2 \cdot 10^{-3}$ | $0.8 \cdot 10^{-3}$ |

**Table 1. Signal efficiency, background rejection and background to signal ratio obtained for events with at least 80% of centre of mass energy given for the centre of mass energy of 500GeV and 1 TeV.**

As it can be seen in Table 1, physics background, taken as the full size effect, contributes to the total systematic error of the integral luminosity no more then few permille. Background to signal ratio is estimated with an approximate statistical error less than 10%.

The inclusion of energy balance criterion of $E_{bal}<0.1\ E_{min}$ would improve the B/S ratio to $1.8 \cdot 10^{-3}$ and $0.7 \cdot 10^{-3}$ for 500 GeV and 1 TeV case, respectively.

Selection criteria based on topology or energy deposition of Bhabha events are correlated to each other, covering the same region of the phase space to large extent. Thus the overall performance of a single criterion on background reduction, results in the same order of magnitude of background to signal ratio, being of the permille order [12].

The inclusion of the acolinearity would improve the B/S ratio by another 10%(20%) for 500GeV and 1 TeV case, respectively. However, the method for correction of beam induced effect rules-out acolinearity criterion [9]. It has been proven [9] that the acoplanarity requirement improve the fraction of the lost events from the luminometer fiducial volume due to the beam induced effects with respect to the case when a selection is based on the deposited energy solely.

## 7. Conclusion

Background from four-fermion processes represents the main source of systematic uncertainty in luminosity measurement at ILC. It contributes to the relative systematic error of the measured luminosity at the level of a few permille at ILC energies.

Having in mind other sources of systematic uncertainty (i.e. the beam induced effects) optimization of the selection criteria resulted in the selection based on the fraction of the center-of-mass energy and maximal allowed acoplanarity of the final state particles. The background to signal separation based on the full set of LEP-type isolation cuts would give slightly better results in the reduction of physics background. However, it has been shown [9] that acolinearity criteria cannot be incorporated in the proposed corrective method for the beam induced effects.

It has to be kept in mind that physics background should be taken as a correction to the Bhabha count once NLO theoretical correction of the four-fermion production cross-section is known at the ILC energies.